# PoEmotion: Can AI Utilize Chinese Calligraphy to Express Emotion from Poems?

[1]Tiancheng LIU, [2]Anqi WANG, [3]Xinda CHEN, [4]Jing YAN, [5]Yin LI, [6]Pan HUI, [7]Kang ZHANG

[1,4,6,7]The Hong Kong University of Science and Technology (Guangzhou) (Guangzhou, China)
[3,5] China Academy of Art (Hangzhou, China)
[2,5,6]The Hong Kong University of Science and Technology (Hong Kong, China)

[1]tcliu767@connect.hkust-gz.edu.cn, [2]awangan@connect.ust.hk, [3]sund971120@gmail.com, [4]jyan856@connect.hkust-gz.edu.cn, [5]ly980810@163.com, [6]panhui@ust.hk, [7]kzhangcma@hkust-gz.edu.cn

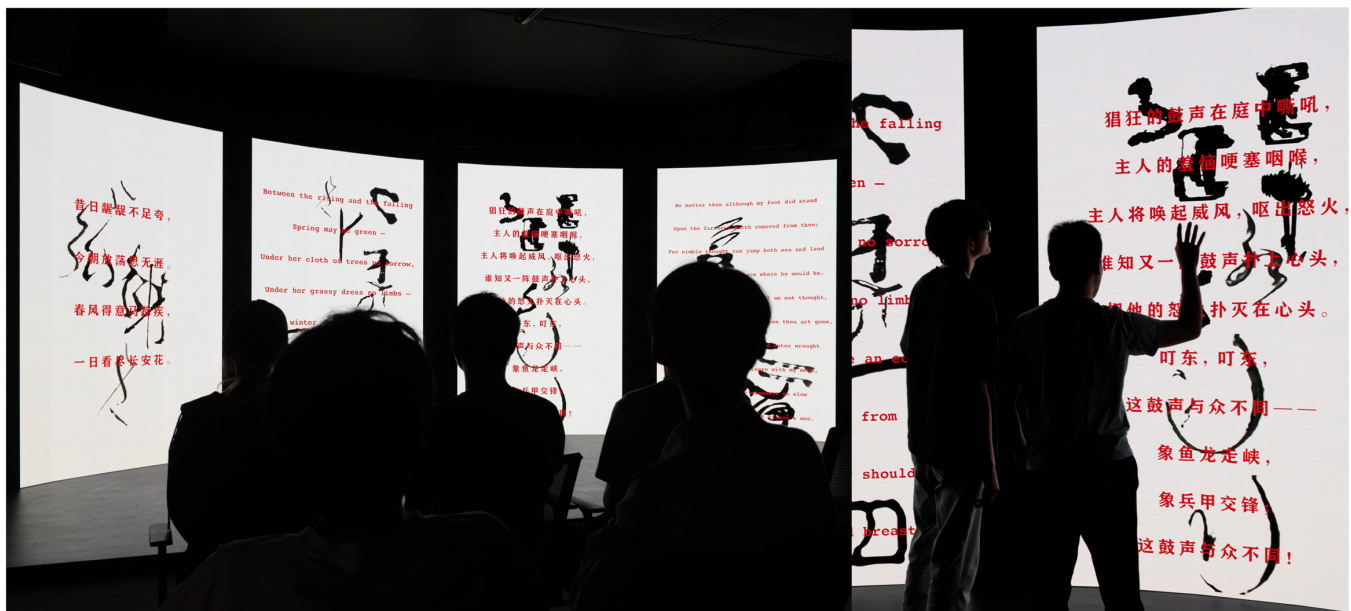

Figure 1. PoEmotion Exhibition. ©Tiancheng LIU, 2023

## Abstract

This paper presents PoEmotion, an approach to visualizing emotions in poetry with Chinese calligraphy strokes. Traditional textual emotion analysis often lacks emotional resonance due to its mechanical nature. PoEmotion combines natural language processing with deep learning generative algorithms to create Chinese calligraphy that effectively conveys the emotions in poetry. The created calligraphy represents four fundamental emotions: excitement, anger, sadness, and relaxation, making the visual representation of emotions intuitive and concise. Furthermore, the approach delves into the relationship between time, emotion, and cultural communication. Its goal is to provide a more natural means of communicating emotions through non-verbal mediums to enhance human emotional expression.

## Keywords

Poem, emotion, visualization, Chinese Calligraphy, NLP, generative art

## Introduction

This paper presents the technical pipeline of PoEmotion (Figure 1) that combines poetry, Chinese calligraphy, natural language processing, and deep learning generative models. We use algorithmically generated non-realistic calligraphic strokes to express the emotions of computed poems. We discuss how readers can feel the emotion of poetry without knowing the cultural background in the modern information society; how poetry intertwined with ancient Chinese calligraphy and future artificial intelligence can resonate with the human political culture.

The emotions find a rich and profound expression in poetry. However, the conveyed emotion in poems to readers is often limited and insufficient when relying solely on plain text. Poets often encapsulate complex emotions within succinct verses, requiring readers to employ their imagination to discern the subtleties. Furthermore, a communication gap exists between the poet's intention and readers, who may lack an understanding of the poet's background and cultural context. This gap results in a mismatch in emotional resonance for poetry.

While Western culture may perceive calligraphy as a design element that revolves around visual beauty. In Chinese culture, calligraphy is an art form reflecting the emotions and philosophical thinking of the writer on a deeper level [1]. Ancient Chinese calligraphers conveyed their emotions and thoughts through various elements, such as the formation of the brush strokes, the intensity of the ink, the structure of the characters, and the overall layout of the artwork [2][3].

PoEmotion seeks to overcome the limitation of conveying emotions through digitally typed text, utilizing artistic visualization. It combines poem's emotion analysis and generative Chinese calligraphy strokes to visually represent the emotion in poems. The emotional scores are integrated with calligraphic strokes that represent basic emotions like excitement, sadness, anger, and relaxation. Each stroke functions as an annotation, visualizing the emotional subtleties in poems. Specifically, the attached emotional strokes in the poetry enhance visual tension and emotional depth, previously lacking in digitally typed words.

Apart from the visualization of emotion in poems, PoEmotion also shows the aesthetic intertwining of time and emotion by combining ancient Chinese calligraphy's unknown and uncontrollable strokes with algorithmically calculated emotional scores. Its endeavor transcends temporal boundaries, creating an emotional resonance for the reader.

Emotion in poems is not just individual feeling. It often touches on common human emotional experiences. Its visualization in calligraphy is also a way of cultural transmission and sharing. As part of the culture, poetry carries the history and memories and conveys emotions through text. The article is structured as follows: Section 2 offers our insight into the connection between poems, emotions, Chinese calligraphy, and visualization. Section 3 delves into the approach's design technology, while Section 4 provides a detailed discussion. Finally, Section 5 concludes the paper.

## Related Work

Poetry is a unique way for people to express their inner world[4]. Traditionally, it carries emotions and far-reaching meanings. The text in poetry conveys the information, touches the reader's heart, and inspires empathy. However, there is an essential challenge to the transmission of emotion. Even the most carefully crafted verses may be understood differently depending on the reader's experience, cultural background, emotional state, and other factors. Poetry's emotional delivery often requires the reader to use imagination and empathy to understand the emotions conveyed through words. This can be a significant emotional effort on the reader's part.

This difficulty in emotional transmission is particularly acute in contemporary society. In the wave of digitization and informatization, the speed and way of reading have changed, and the fast-paced life and segmentation of information make it increasingly difficult for people to immerse themselves in the emotional world of poetry. Therefore, how to effectively convey emotions through poetry in modern society has become a problem to be solved.

Emotions and their categorization have been one of the core areas of psychological research. Scholars have proposed various emotion models to describe and categorize human emotions. Some of the well-known models include Plutchik's wheel model of emotion, which proposes a multidimensional perspective of emotion [5]. Ekman's grounded emotion theory emphasizes the universality of emotional expression in global cultures [6]. Mehrabian's and Russell's PAD model of emotion states proposes three main dimensions of emotional experience [7]. Russell's Circumplex Model of Emotion emphasizes the bi-dimensional nature of emotion [8]. James' attribution of emotion to levels of bodily involvement comprises of four basic emotions [9]. Lazarus and Lazarus expand the list of emotions to fifteen [10]. Cowen and Keltner's in-depth study and data analysis has identified twenty-seven different categories of emotions [11].

Emotion Computing, as an essential branch of Artificial Intelligence, builds systems to respond and adapt to human emotion [12][13]. Research in this field focuses on recognizing the emotional aspects of various signals, such as text, speech, expressions, and sounds, to convert them into measurable data.

In text emotion analysis, traditional methods like TF-IDF have been widely applied [14]. With the advent of deep learning, Recurrent Neural Networks (RNNs) [15] and long short-term memory networks (LSTMs) [16], [17] have achieved higher precision in emotion classification tasks. Notably, benefiting from the Attention Mechanism, the Transformer model can batch process all data points in a sequence, significantly improving the training efficiency over traditional RNNs and LSTMs [18]. The Transformer-based BERT model is frequently used in sentiment analysis classification tasks [12] [13], and there is research utilizing BERT for detecting emotions in poetry [21]. It gets a higher classification result. So, the BERT model is employed in PoEmotion.

Kucher et al. analyzed 132 scientific visualization techniques for sentiment detection in text data [22]. Research in visualizing textual emotions is extensive, yet studies focusing on expressing these emotions through text are limited. TextTone developed an online chat software that changes fonts, sizes, and colors to reflect different emotions. However, this approach can compromise readability and may not always effectively convey the intended emotions.

FaceType focuses on the direct interaction between spoken and written forms, instantly converting verbal expressions into Chinese calligraphic text [23]. Therefore, this approach proposes a new visualization medium - calligraphic strokes - aiming to break the boundaries of traditional visualization through the dynamics and aesthetics of calligraphy and enable the audience to visualize the emotions behind the text. But this work uses sound only as an emotional input, without considering the text. Two other works analyze the text's emotion to generate the appropriate calligraphy [24], [25], making their strokes and layouts different. However, for non-Chinese texts, their works are unable to generate Chinese calligraphy that expresses emotions.

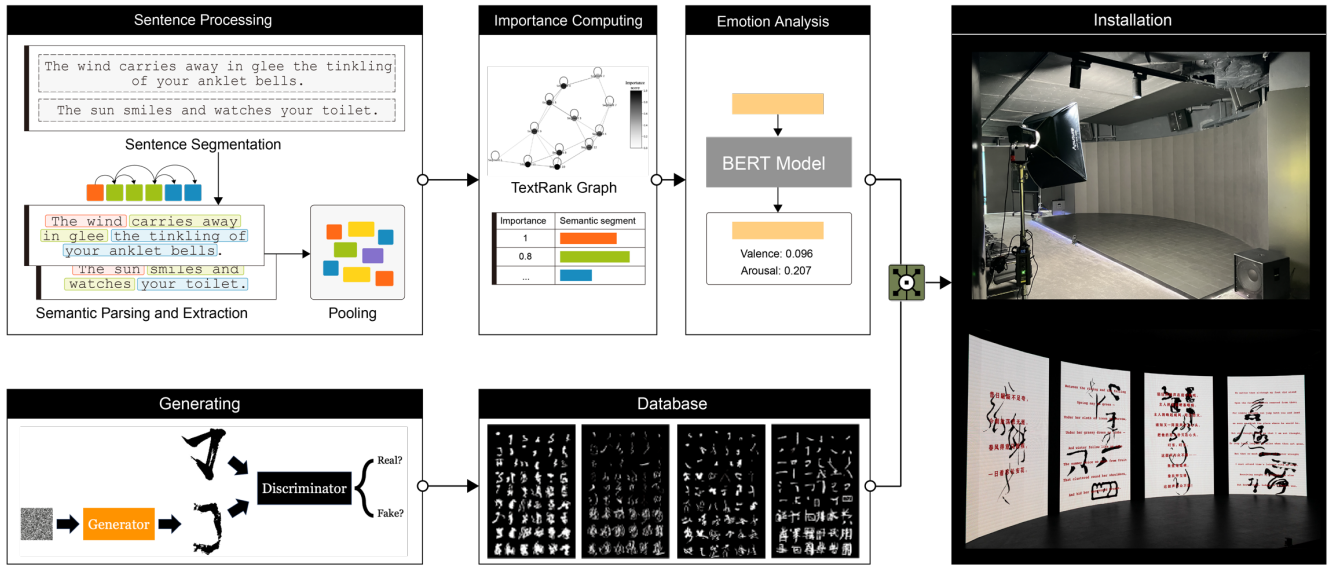
Figure 2. PoEmotion Framework

## PoEmotion

PoEmotion comprises three main parts: emotional analysis of poem text, generation of Chinese calligraphy strokes, and integration and presentation of the work (Figure 2), as detailed below:

- **Emotional Analysis of Poem Text:**
   1. Process poem text for sentence segmentation.
   2. Perform semantic parsing and extraction on each sentence.
   3. Aggregate semantic segments into a pool ranked on their emotional significance.
   4. Conduct emotion analysis on top-ranked segments for classification and scoring.
- **Generation of Calligraphy Strokes:**
   1. Train a Generator using Trans GAN and a database of real segmented strokes.
   2. Generate strokes that represent various emotions.
- **Integration and Presentation:**
   1. Pair emotional scores from semantic segments with corresponding strokes.
   2. Integrate these elements for display.

We aim to present the selected semantic segments alongside their corresponding strokes. This synthesis of text and image provides a multidimensional perspective on the emotional fabric of the poem, offering both quantitative and qualitative insights.

### Sentence Segmentation

For sentence segmentation, we employ the Natural Language Toolkit (NLTK) to subdivide the text into independent and meaningful sentence units [26]. This process identifies and separates sentences to provide a clear basis for subsequent text analysis and language processing.

The Punkt algorithm in NLTK segments sentences using a rule and context-based approach for determining sentence boundaries. First, it relies on end-of-sentence punctuation marks, such as periods, question marks, and exclamation points. It segments the text into candidate sentences and applies a set of syntactic and semantic rules to further filter and adjust these candidate sentences to generate an accurate list of sentences ultimately. These rules consider the presence of punctuation and the context of the punctuation and other factors related to syntax and semantics. As a result, the Punkt algorithm can recognize the end of a sentence in a text and classify the text into semantically coherent sentence-level units.

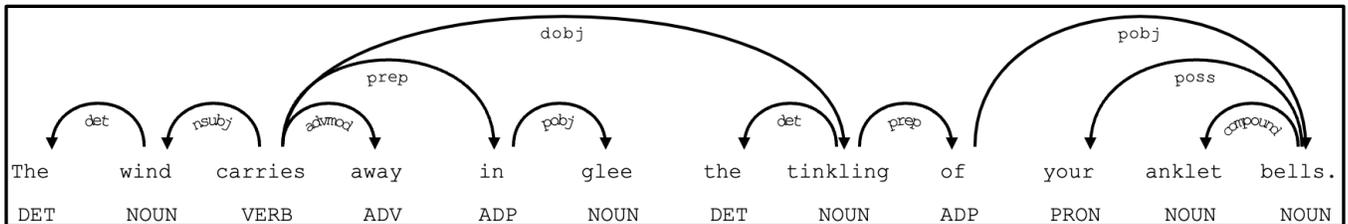
Figure 3. Semantic Parsing Dependency Tree

### Semantics Paring and Segmentation

We use the spaCy Transformers model for semantic parsing [27], of the components of a sentence, extract key linguistic features, and encode them into a vector form (Figure 3).

We implement semantic parsing by means of dependency syntactic analysis provided by spaCy. Dependency syntactic analysis helps to understand the dependencies between words in a sentence and thus captures the semantic structure at a finer granularity. By analyzing dependency trees, Algorithm 1 can identify groups of words that are semantically closely related. We extract noun phrases (including their modifiers) and verb phrases (including verbs and their direct objects and gerunds) as semantic segments.

---

**Algorithm 1** Sentence Processing Logic

1: Process Document to obtain syntax structure
2: **for** each sentence in the document **do**
3:     Find Subject with 'nsubj' dependency
4:     Find Verb Phrases with 'ROOT' and 'conj'
5:     Check for Verb Objects with 'dobj' depende

---

### Semantic Segments Pooling

We take the segmented semantic segments from all the above sentences and put them into a pool. This makes it easier for the subsequent processes.

### Semantic Segments Importance Scoring and Ranking

This process is divided into constructing semantic segment graphs, computing importance scores, ranking, and selection.

**Constructing semantic segment graphs:** We construct a graph using the TextRank model [20]. TextRank is an algorithm that ranks nodes in the graph and each semantic segment is considered a node in the graph. All semantic segments are first represented as nodes, and edges in the graph are built based on their semantic similarity using Word Embeddings. These edges represent the relationships between different semantic segments.

**Computing Importance Score:** The TextRank algorithm considers the relationship between nodes and their neighboring nodes by iteratively computing the weights between nodes. It calculates the importance score of each semantic segment in the whole corpus. Higher scores are given to semantic segments that are more important in the overall emotion structure.

**Sorting and Selection:** The computed importance scores are used to sort the semantic segments in descending order. Typically, the option is to keep the top 50%, or some other percentage of high-importance semantic segments, and filter out other less important segments. This helps to reduce the computational complexity of subsequent emotion analysis by focusing on the parts of the text related to emotion.

### Semantic Segments Emotion Analysis

We adopt Russell's model of emotion as the theoretical basis for emotion analysis [8]. The model distributes emotions on a plane consisting of two orthogonal dimensions: Valence and Arousal. The valence dimension reflects the positive or negative direction of emotion, while the Arousal dimension depicts the energy level of emotion, which varies from a low-energy (e.g., relaxation or sadness) to a high-energy (e.g., anger or excitement) state.

We use the Bidirectional Encoder Representations from Transformers (BERT) model for sentiment analysis. BERT is a state-of-the-art natural language processing technique that understands the context of a text and extracts its deep semantic messages. We use a pre-trained `bert-base-uncased` model to predict the valence and arousal scores of the sentiment in the text, which reflect the two main dimensions of the sentiment in the text. This process provides an analytical foundation for sentiment visualization.

Finally, we use the Euclidean Distance where Valence and Arousal are located for the computation of emotion intensity.

$$Intensity_{emotion} = \sqrt{Score_{Valence}^2 + Score_{Arousal}^2}$$

Combining these approaches, our analysis produces two types of output: the type of emotion, such as excitement, sadness, anger, or relaxation, and an intensity label for the corresponding emotion type. Figure 4 shows an example of emotion classification results. This enables us to perform multidimensional quantitative analysis of complex emotional expressions in text samples. This analysis enhances our understanding of how language conveys emotional meanings and provides possible insights into the interplay between affective computation and visual presentation of textual information.

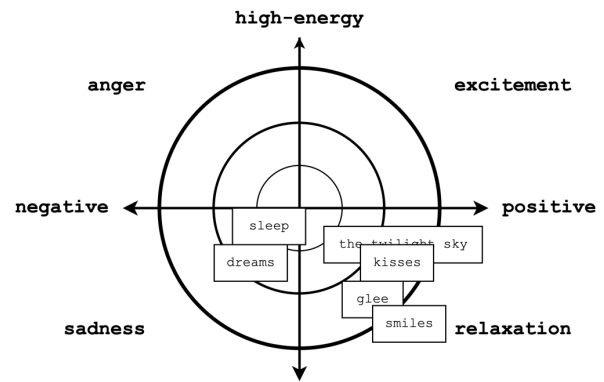

Figure 4. Example Classification in Emotion Model

## Calligraphy and Emotion

Russell's emotion model's concise and intuitive categorization makes it easy to translate our complex emotional experiences into recognizable emotional expressions in Chinese calligraphy.

Our approach corresponds the brush strokes of calligraphy to the emotional quadrants, making the visual expression of emotions intuitive and concise. Through the visualization of emotions in Chinese calligraphy, an ancient art form, we can provide people with a non-verbal way of communicating emotions, thus enriching the level and depth of human emotional expression.

Chinese calligraphy's strokes and lines are a carrier of text and an important tool for emotional expression. Every brushstroke of calligraphers contains his/her emotional state and spiritual mood, like silent music, conveying rich emotions through the brush's thickness and speed, and the continuity and breakage of strokes. The following section discusses the emotional interpretations of the works of four calligraphers. Through their strokes, we can glimpse the subtle connection between calligraphy and emotion.

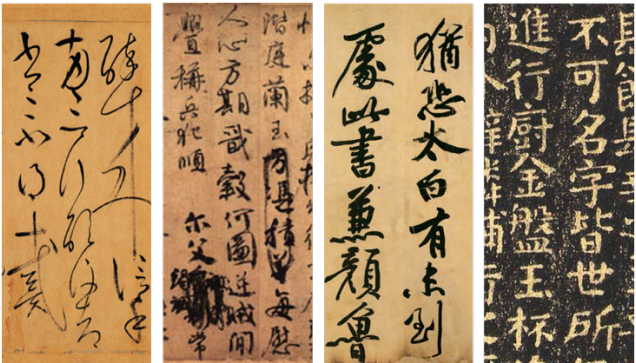

Figure 5. part view of the Chinese calligraphy artwork
《自叙帖》,《祭侄文稿》,《题苏轼寒食帖跋》,《麻姑山仙坛记》

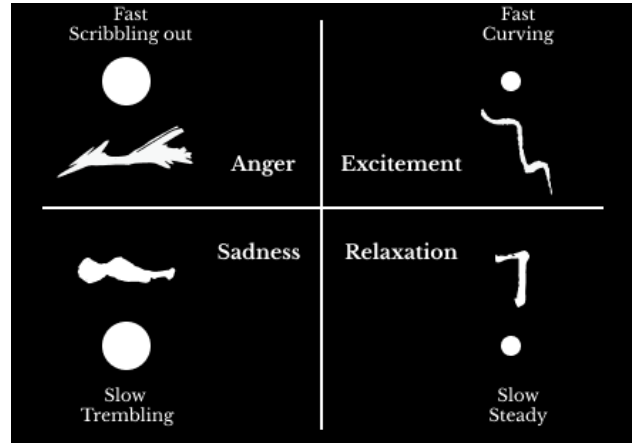

Figure 6. Stroke Features in Different Emotion

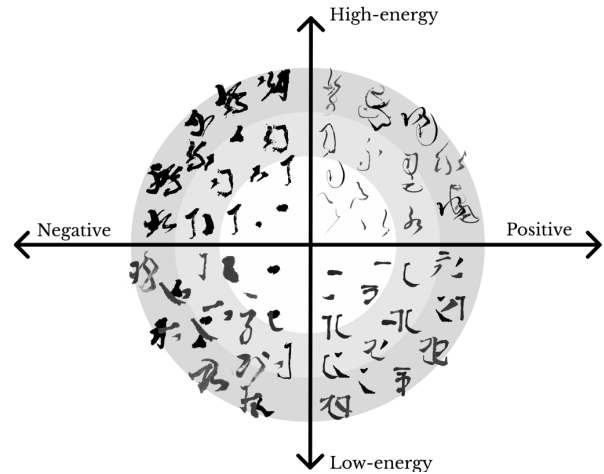

Figure 7. Part of the Emotion in Valence-Arousal Space

**Excitement:** ZHANG Xu's *self-narrative post* (《自叙帖》, first from the left in Figure 5) is famous for its wild cursive style, where the lines and strokes are full of tension and indulgence as if dancing wildly on the paper [28]. The speed and fluidity of the strokes and the character shapes' expansion reveal uncontrollable excitement and enthusiasm. Behind this excitement is Zhang Xu's fervor for life and his strong expression of self. His strokes are turbulent and uninhibited, like his emotions, pouring his agitated inner world onto the paper.

**Anger:** YAN Zhenqing's *Manuscript of Offering to a Nephew* (《祭侄文稿》, 2nd from the left) expresses the calligrapher's anger and grief with his bold and powerful strokes [29]. The heavy force of the strokes and the sharpness of the brush reveal Yan Zhenqing's inner rage and indignation. When composing this manuscript, Yan Zhenqing experienced a personal tragedy - the untimely death of his nephew being framed due to political struggles. His strong emotions are conveyed through his brushstrokes so that future generations can feel the extremely angry emotional fluctuations behind the text.

**Sadness:** In HUANG Tingjian's *Inscription on SU Shi's Cold Food Post* (《题苏轼寒食帖跋》, 3rd from the left), the trembling and pauses of the brushstrokes silently tell of the feelings and sadness that follow the reading [30]. The trembling of the brushstroke is like the fluctuation of the heart, and the technique of trembling the brush aptly emphasizes HUANG Tingjian's deep sorrow in the face of his friend SU Shi's text and the delicacy and depth of this emotion is subtly conveyed in the brushstrokes.

**Relaxation:** YAN Zhenqing's *Record of the Immortal Altar on Mount Magu* (《麻姑山仙坛记》), on the other hand, presents a mood of serenity and contentment [31]. The strokes are steady and smooth, the lines are graceful, with a few drastic changes. This style of writing reflects Yan Zhenqing's appreciation of the natural beauty of Mount Magu and the mental relaxation and pleasure gained from being in nature. The calligrapher's brushstrokes depict mythological stories with an emotional touch of relaxation.

Through these four examples, we can see that a calligrapher's emotions are expressed through the twists and turns of the brush, the changes in the brush's strength, and the rhythm's speed. Here, excitement is reflected in the brush stroke of fast and light curve; anger in that of fast, heavy, scribbling out; sadness in that of slow, heavy, trembling; and relaxation in that of light, slow, and stable.

PoEmotion explores this unique way of emotional expression and makes these traditional emotional symbols vividly presented using modern technology.

## Stroke Generator and Stroke Database

Our stroke generator employs the TransGAN model and stroke database [32]. The model generates calligraphic strokes that reflect specific emotions. We select the four famous aforementioned calligraphy works with strokes segmented as the training set to achieve this. Subsequently, these segmented strokes are assigned to four different Trans GAN models, each focusing on learning and modeling the stroke style of a calligrapher.

The Trans GAN model is developed from the GAN model [25]. The GAN model comprises a Generator (G) and a Discriminator (D). The Generator generates visually convincing stroke images, while the Discriminator determines whether the images are close to real calligraphy strokes. The generator and discriminator compete during the training process and eventually reach a balance where the generator can create more and more realistic strokes, and the discriminator is able to tell the difference between the generated image and real stroke with increasing accuracy.

The model uses the Transformer architecture for both the generator and discriminator to capture calligraphic strokes fluently.

The training process is divided into two phases. First, the parameters of the generator are fixed, random noise is fed into the generator to produce image. Then these generated images are fed into the discriminator along with images of real strokes. In this step, the parameters of the discriminator are updated to recognize the real images as positive samples (labeled 1) and the generated images as negative samples (labeled 0) . In the second phase, the parameters of the discriminator are fixed, and updated based on the feedback from the discriminator to produce a more realistic image, causing the discriminator with fixed parameters to misidentify the generated image as the real image (labeled 1 whenever possible). By iterating these two phases, the model is gradually optimized until the generator can produce highly realistic stroke images. Overall, the philosophy is concluded in below:

$$\min_{G} \max_{D} V(D, G)$$
$$V(D,G) = \mathbb{E}_{x \sim p_{data}(X)}[\log D(x)] + \mathbb{E}_{z \sim p_z(z)}[\log(1 - D(G(z)))]$$

where:
- $G(z)$ is Generated Data,
- $D(x)$ is the Probability Value (ranging from 0 to 1),
- $\mathbb{E}$ denotes the Expectation,
- $p_{data}$ represents the Distribution of Real Data,
- $p_z$ is the Distribution of the Input Noise to the Generator.

Perceived as intense emotions, each complex stroke is calculated with its complexity:

$$\text{Complexity} = \frac{\text{Perimeter}^2}{\text{Area}}$$

where:
- *Complexity* is the complexity of the calculated stroke,
- *Perimeter* is the perimeter of the stroke's contour,
- *Area* is the area of the stroke [33].

Finally, the emotional analysis results of semantic segments are matched with strokes in the stroke database, as shown in Figure 7. The matching strokes based on the emotion scores of the semantic segments sent to the output.

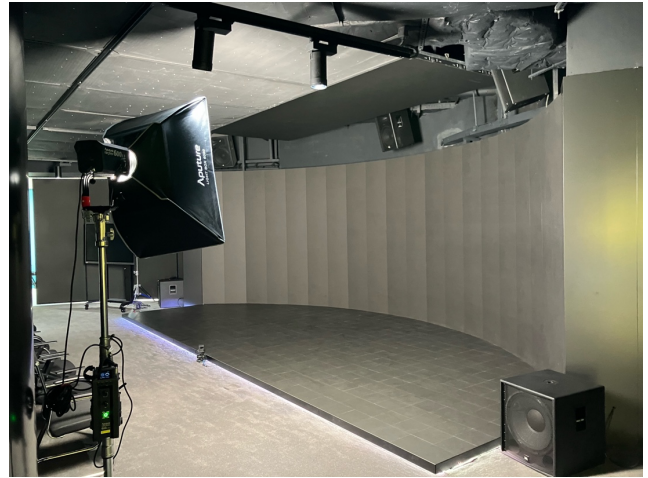

Figure 7. Display Space

## Display:

The PoEmotion installation showcases poems accompanied by their corresponding generative calligraphy strokes. They are displayed on a large, curved screen with a resolution of 5120x1536, optimized for an HDMI output of 3840x1152. The display is integrated with Touch Designer software, which facilitates the dynamic presentation of the calligraphic strokes alongside poems.

## Discussion

The initial intention of PoEmotion is to explore a core issue: how to complement text in emotional transmission with artistic visualization. In the emotional analysis and visual presentation of poetry, we attempt to capture those delicate layers of emotion and emotional colors that are difficult to express with pure text. By introducing the visual elements of strokes, we analyze and quantify poems' emotions, with an artistic interpretation of the beauty of the intertwining of time and emotion.

PoEmotion finds that although poetry is rich in emotional expression, certain emotions are hard to convey. The generative strokes can abstractly reflect the emotional texture of the poem, providing a new perceptual dimension for the audience. Embedding the temporal level of emotion in the poem stereoscopic, visual presentation allows readers to intuitively experience the depth of emotion that the poet is trying to convey.

PoEmotion also reveals the complex relationship between text and emotional expression. Poetry's emotional essence is difficult to quantify and visualize due to its close connection to personal experiences and cultural contexts, demanding individuality and emotional diversity. By constructing a coordinate system of emotions and correlating it with a database of strokes, PoEmotion reveals the interplay between individual emotions and collective memory. Different individuals may perceive and express the same emotion differently. Therefore, our approach must be flexible enough to address users' emotional experiences and cultural differences.

To summarize our findings, this study provides an insightful perspective that the artistic expression of emotion is not just a one-way transmission but an interactive and empathic process.

Sentiment analysis models may not be able to capture the subtle and complex emotions in poetry with complete accuracy. This limitation may affect the quality of emotional expression in the final generated calligraphic works. People from different cultures may have very different understandings of poetry and calligraphy. Such cultural differences may lead to different interpretations of emotional expressions, which in turn affects the acceptance and understanding of calligraphic works.

The current work focuses on poetry and Chinese calligraphy in English and Chinese. Extending it to other languages and writing systems may require additional adjustments and considerations to ensure accurate communication of emotions. The current approach does not provide opportunities for the audience to directly engage and influence the output, which may limit the interest and engagement of the audience in PoEmotion.

Future research could explore advanced sentiment analysis models to capture the emotional details and complexities more accurately in poetry. It will also explore how PoEmotion can be applied to poetry in different cultural and linguistic contexts to enable wider cultural communication and understanding. Richer interaction mechanisms will also be added to increase user engagement and enhance the educational and entertainment value of the system. Through such efforts, we can expect to uncover more profound insights into the intersection of poetry, art, and emotion, bringing new inspiration and contributions to the field of affective computing.

## Conclusion

This study aims to explore how to compensate for the limitations of text in conveying emotions by translating the emotions of poetic texts into visual strokes. Through a carefully designed process of semantic analysis, emotion scoring, and image generation, we facilitate visualizing the abstract emotions of the poems and give them a new visual life.

We also realize that the subjectivity and complexity of emotion remain a challenge. Future research needs to incorporate more interactions as well as delve into the differences in emotion expression across cultures and individual contexts and how these differences can be accurately reflected in emotion visualization.

We expect this multidimensional approach to emotion parsing will find its value and place in future literary research, emotional computing, and artistic creation.

## Reference

[1] M. H. Müller-Yao, *The influence of Chinese calligraphy on Western Informel painting*. Heidelberg University Library, 2016. doi: 10.11588/ARTDOK.00003932.
[2] X. Shi, 'Chinese Calligraphy as "Force-Form"', *J. Aesthetic Educ.*, vol. 53, no. 3, pp. 54–70, Oct. 2019, doi: 10.5406/jaesteduc.53.3.0054.
[3] X. Shi, 'As If One Witnessed the Creation: Rethinking the Aesthetic Appreciation of Chinese Calligraphy', *Philos. East West*, vol. 70, no. 2, pp. 485–505, 2020, doi: 10.1353/pew.2020.0031.
[4] C. Leggo, 'The heart of pedagogy: on poetic knowing and living', *Teach. Teach.*, vol. 11, no. 5, pp. 439–455, Oct. 2005, doi: 10.1080/13450600500238436.
[5] R. Plutchik, 'A GENERAL PSYCHOEVOLUTIONARY THEORY OF EMOTION', in *Theories of Emotion*, Elsevier, 1980, pp. 3–33. doi: 10.1016/B978-0-12-558701-3.50007-7.
[6] R. W. Levenson, P. Ekman, K. Heider, and W. V. Friesen, 'Emotion and autonomic nervous system activity in the Minangkabau of West Sumatra.', *J. Pers. Soc. Psychol.*, vol. 62, no. 6, pp. 972–988, 1992, doi: 10.1037/0022-3514.62.6.972.
[7] J. A. Russell, L. M. Ward, and G. Pratt, 'Affective Quality Attributed to Environments: A Factor Analytic Study', *Environ. Behav.*, vol. 13, no. 3, pp. 259–288, May 1981, doi: 10.1177/0013916581133001.
[8] J. A. Russell, 'A circumplex model of affect.', *J. Pers. Soc. Psychol.*, vol. 39, no. 6, pp. 1161–1178, Dec. 1980, doi: 10.1037/h0077714.
[9] W. James, 'Discussion: The physical basis of emotion.', *Psychol. Rev.*, vol. 1, no. 5, pp. 516–529, Sep. 1894, doi: 10.1037/h0065078.
[10] R. S. Lazarus and B. N. Lazarus, *Passion and reason: making sense of our emotions*. New York Oxford: Oxford University Press, 1996.
[11] A. S. Cowen and D. Keltner, 'Self-report captures 27 distinct categories of emotion bridged by continuous gradients', *Proc. Natl. Acad. Sci.*, vol. 114, no. 38, Sep. 2017, doi: 10.1073/pnas.1702247114.
[12] R. Calvo, S. D'Mello, J. Gratch, and A. Kappas, Eds., *The Oxford Handbook of Affective Computing*. Oxford University Press, 2015. doi: 10.1093/oxfordhb/9780199942237.001.0001.
[13] C. Marechal *et al.*, 'Survey on AI-Based Multimodal Methods for Emotion Detection', in *High-Performance Modelling and Simulation for Big Data Applications*, vol.

11400, J. Kołodziej and H. González-Vélez, Eds., in Lecture Notes in Computer Science, vol. 11400. , Cham: Springer International Publishing, 2019, pp. 307–324. doi: 10.1007/978-3-030-16272-6_11.

[14] V. Sundaram, S. Ahmed, S. A. Muqtadeer, and R. Ravinder Reddy, 'Emotion Analysis in Text using TF-IDF', in *2021 11th International Conference on Cloud Computing, Data Science & Engineering (Confluence)*, Jan. 2021, pp. 292–297. doi: 10.1109/Confluence51648.2021.9377159.

[15] D. Kher, 'Multi-label emotion classification using machine learning and deep learning methods', Thesis, Laurentian University of Sudbury, 2021. Accessed: Nov. 15, 2023. [Online]. Available: https://zone.biblio.laurentian.ca/jspui/handle/10219/3662

[16] Y. Zhang, J. Zheng, Y. Jiang, G. Huang, and R. Chen, 'A Text Sentiment Classification Modeling Method Based on Coordinated CNN-LSTM-Attention Model', *Chin. J. Electron.*, vol. 28, no. 1, pp. 120–126, 2019, doi: 10.1049/cje.2018.11.004.

[17] J.-H. Wang, T.-W. Liu, X. Luo, and L. Wang, 'An LSTM Approach to Short Text Sentiment Classification with Word Embeddings', in *Proceedings of the 30th Conference on Computational Linguistics and Speech Processing (ROCLING 2018)*, C.-C. (Jeremy) Lee, C.-Z. Yang, and J.-T. Chien, Eds., Hsinchu, Taiwan: The Association for Computational Linguistics and Chinese Language Processing (ACLCLP), Oct. 2018, pp. 214–223. Accessed: Nov. 15, 2023. [Online]. Available: https://aclanthology.org/O18-1021

[18] A. Vaswani *et al.*, 'Attention is All you Need', in *Advances in Neural Information Processing Systems*, I. Guyon, U. V. Luxburg, S. Bengio, H. Wallach, R. Fergus, S. Vishwanathan, and R. Garnett, Eds., Curran Associates, Inc., 2017. [Online]. Available: https://proceedings.neurips.cc/paper_files/paper/2017/file/3f5ee243547dee91fbd053c1c4a845aa-Paper.pdf

[19] Z. Gao, A. Feng, X. Song, and X. Wu, 'Target-Dependent Sentiment Classification With BERT', *IEEE Access*, vol. 7, pp. 154290–154299, 2019, doi: 10.1109/ACCESS.2019.2946594.

[20] M. Munikar, S. Shakya, and A. Shrestha, 'Fine-grained Sentiment Classification using BERT', in *2019 Artificial Intelligence for Transforming Business and Society (AITB)*, Nov. 2019, pp. 1–5. doi: 10.1109/AITB48515.2019.8947435.

[21] T. Haider, S. Eger, E. Kim, R. Klinger, and W. Menninghaus, 'PO-EMO: Conceptualization, Annotation, and Modeling of Aesthetic Emotions in German and English Poetry'. arXiv, Jun. 23, 2021. Accessed: Nov. 15, 2023. [Online]. Available: http://arxiv.org/abs/2003.07723

[22] K. Kucher, C. Paradis, and A. Kerren, 'The State of the Art in Sentiment Visualization', *Comput. Graph. Forum*, vol. 37, no. 1, pp. 71–96, 2018, doi: 10.1111/cgf.13217.

[23] K. Maher, F. Xiang, and L. Zhi, 'FaceType: Crafting Written Impressions of Spoken Expression', *Proc. ACM Comput. Graph. Interact. Tech.*, vol. 5, no. 4, pp. 1–9, Sep. 2022, doi: 10.1145/3533385.

[24] 'Learning to Compose Stylistic Calligraphy Artwork with Emotions | Proceedings of the 29th ACM International Conference on Multimedia'. Accessed: Nov. 15, 2023. [Online]. Available: https://dl.acm.org/doi/10.1145/3474085.3475711

[25] R. Liu, S. Yuan, M. Chen, B. Chen, Z. Qiu, and X. He, 'MaLiang: An Emotion-driven Chinese Calligraphy Artwork Composition System', in *Proceedings of the 28th ACM International Conference on Multimedia*, in MM '20. New York, NY, USA: Association for Computing Machinery, Oct. 2020, pp. 4394–4396. doi: 10.1145/3394171.3416338.

[26] S. Bird, E. Klein, and E. Loper, *Natural language processing with Python*, 1st ed. Beijing ; Cambridge [Mass.]: O'Reilly, 2009.

[27] Ines Montani, Matthew Honnibal, Matthew Honnibal, Adriane Boyd, Sofie Van Landeghem, and Henning Peters, 'explosion/spaCy: v3.7.2: Fixes for APIs and requirements'. Zenodo, Oct. 16, 2023. doi: 10.5281/ZENODO.1212303.

[28] Q. Zhang, 'The Treasure House of Ancient Chinese Literature and Art', in *An Introduction to Chinese History and Culture*, in China Academic Library. , Berlin, Heidelberg: Springer Berlin Heidelberg, 2015, pp. 319–351. doi: 10.1007/978-3-662-46482-3_11.

[29] S. Wang, 'The Upright Brush: Yan Zhenqing's Calligraphy and Song Literati Politics (review)', *China Rev. Int.*, vol. 9, no. 1, pp. 191–196, 2003, doi: 10.1353/cri.2003.0064.

[30] P. De Vries, 'Passing the Shrine of the God Calming the Waves and the Notion of Emptiness in Huang Tingjian's (1045–1105) Calligraphy', 2010, doi: 10.5167/UZH-42511.

[31] L. Yang, 'Images for the Temple: Imperial Patronage in the Development of Tang Daoist Art', *Artibus Asiae*, vol. 61, no. 2, p. 189, 2001, doi: 10.2307/3249910.

[32] Y. Jiang, S. Chang, and Z. Wang, 'TransGAN: Two Pure Transformers Can Make One Strong GAN, and That Can Scale Up'. arXiv, Dec. 08, 2021. Accessed: Nov. 09, 2023. [Online]. Available: http://arxiv.org/abs/2102.07074

[33] J. E. Cutting and J. J. Garvin, 'Fractal curves and complexity', *Percept. Psychophys.*, vol. 42, no. 4, pp. 365–370, Jul. 1987, doi: 10.3758/BF03203093.